# High-SNR snapshot multiplex spectrometer with sub-Hadamard-S matrix coding

Zhuang Zhao[1], Lianfa Bai[1], Jing Han[1], Jiang Yue[2*]
1 School of Electronic Engineering and Optoelectronic Technology, Nanjing University of Science and Technology, Nanjing 210094, China
2 National Key Laboratory of Transient Physics, Nanjing University of Science and Technology, Nanjing 210094, China
* 190281182@qq.com

**Abstract** We present a robust high signal-to-noise ratio (SNR) snapshot multiplex spectrometer with sub-Hadamard-S matrix coding. We demonstrated for the first time that the sub-Hadamard-S matrix coding could provide comparable SNR improvement with Hadamard-S matrix in Hadamard transform spectrometer (HTS). Normally, HTS should change the coding mask to obtain a reasonable spectrum result, causing unexpected time-consuming. An extra imaging path to collect the light intensity of the aperture is added in this paper. Both light intensity of the aperture and overlapped spectra are captured within one shot, turning Hadamard-S matrix coding into sub-Hadamard-S matrix coding. Simulations and experiments show that the proposed method could obtain comparable SNR improvement with the traditional HTS, maintaining snapshot.

## 1. Introduction

Spectrometer is an important tool used in widely scientific research, such as, chemical detection, medicine research. There are two fundamental problems challenge the result accuracy of the spectrometer, one is the weak source, the other is transient phenomenon. Sometimes it should deal with both of the problems, for example, the research of laser-matter interaction. There are two mainly remarkable technologies are proposed to address the spectrum measurement of weak source, Hadamard Transform Spectroscopy (HTS) and Fourier Transform Spectroscopy (FTS). Unfortunately, both of the technologies need many extra measurements to improve the SNR of result, which means that both of them can't deal with spectrum of weak source in transient phenomena. Snapshot measurement maintaining high signal-to-noise ratio (SNR) of result is demanded. Besides, as a measurement reliable improvement is very important, which means it should have a certain lower bound.

In order to realize high throuput advantage within snapshot, some creative ideas have been proposed, such as, spectrometer based on compressive sensing [1-6], computational slits [7, 8] and deconvolution [9]. These methods try to improve the SNR and resolution of spectrum with numerical methods. However, compared with the coding aperture, the numerical methods are limited by the algorithm and structure of signal. Specifically, the retrieval algorithms are proposed to deal with the so-called ill-posed inverse problems, which means it cannot guarantee certain and robust SNR improvement. The estimated results depend on the structure of signal and type of noise. The traditional multiplexing methods, such as HTS [10-12], guarantee certain improvement of SNR in detector noise dominated. Especially, the SNR boost would increase along with the scale of coding matrix [13-15]. Unfortunately, the traditional multiplexing would cause unexpected heavy time consumption. Spectrometer with certain and continuous SNR boost, maintaining snapshot, would enormously extend the application.

Among difference types of coding matric, Hadamard-S matrix is well-structured and easily to implement. Besides, it is the mostly used coding matrix in multiplexing, based on which a static, multimodal, multiplex spectrometer (MMS) is proposed [16]. In MMS, based on compressed sensing (CS), Nonnegative least-squares (NNLS) [16-18] method is employed to reconstruct the spectrum of the scene. It could realize a snapshot measurement keeping high-throughput advantage, and even estimate spectral images of target. MMS has a compact optical struct and can obtain high SNR measured results, however, in MMS the light intensity is represented by coding matrix and make the reconstructed results unstable. A dual-camera coded aperture snapshot spectral imaging is proposed [19] to overcome this drawback. The authors add an extra imaging path to detect the light intensity and then use two-step iterative shrinkage/thresholding (TwIST) [20] to obtain the final results. However, according to the theory of CS, the accuracy of reconstructed result depends on the structure of signal.

In this paper, we introduce a new coding matrix called sub-Hadamard-S matrix, which gains HTS comparable SNR improvement, maintaining snapshot. The proposed method contains two imaging paths: non-dispersive imaging path and dispersive imaging path. The non-dispersive imaging path is employed to measure the light intensity at the coding aperture. The dispersive imaging path is employed to measure the overlapped dispersive spectra. Due to the variation of the light intensity at the aperture, the standard Hadamard-S matrix coding would be turned into a sub-Hadamard-S matrix. We demonstrated that the sub-Hadamard-S matrix coding could realize a proximity denoising capability with the Hadamard-S matrix coding. Most important, it could provide a tool to address the problem of spectrum detection of transient weak signal.

## 2. Implementation of snapshot multiplex spectrometer

The implementation and scheme of the proposed snapshot spectrometer with sub-Hadamard-S matrix coding are shown in Fig. 1(a) and 1(b), respectively. In the implementation, Hadamard-S matrix is employed to code the light at aperture. A beam splitter (split ratio, 70:30) is utilized to split the beam into two. One is reformed to detect the light intensity distribution of the aperture, which is employed to retrieve the coding matrix, called non-dispersive path. It would be a sub-Hadamard-S matrix, as the light intensity of pixels would not be the same as each other. Another light beam is dispersed by the blazed grating, and is reformed as overlapped spectra of different pixels at aperture. The system contains two CCDs and they work at the same time and controlled through the synchronizing signal of DMD. The $CCD_1$ is employed to collect the overlapped dispersed spectra and $CCD_2$ is employed to collect the non-dispersed image of the aperture on DMD. The ground glass is employed to uniform the light intensity. The DMD used in this paper is D-ED01N, its resolution is 1024×768 and the pixel size is 13.68μm×13.68μm. The CCD is Basler acA1920-155um, its resolution is 1920×1200 and the pixel size is 5.86μm×5.86μm.

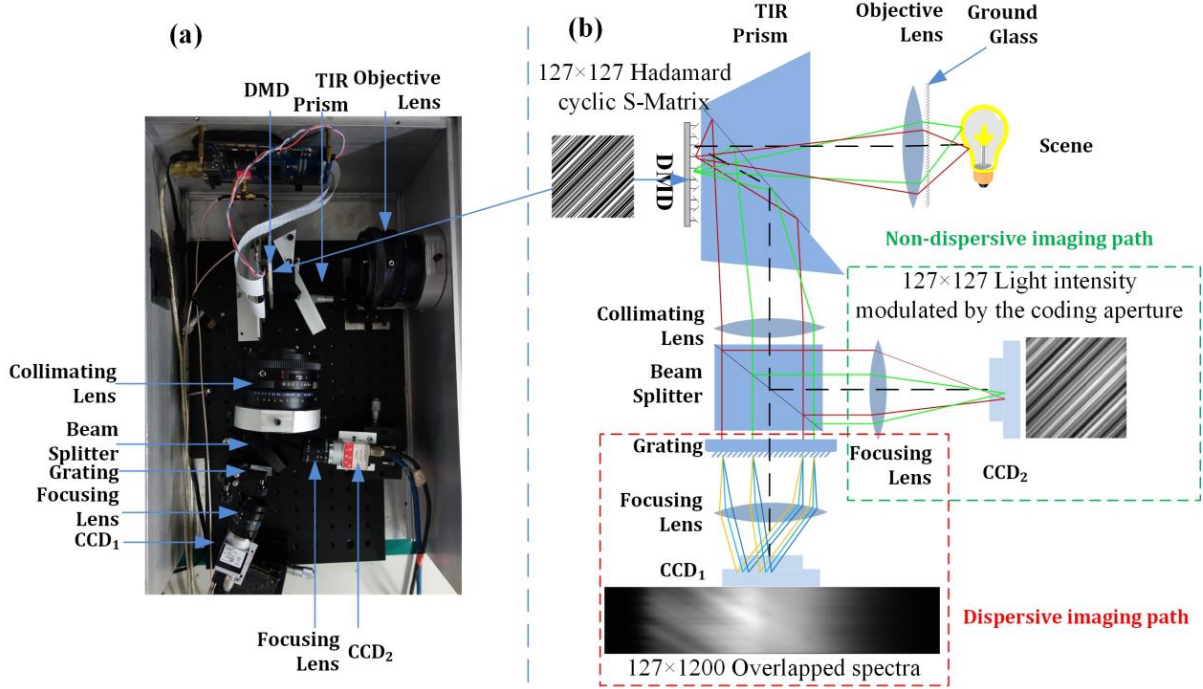

Fig.1. Overview of the snapshot multiplexing spectrometer. (a) The implementation of the system. (b) Schematic drawing of the implementation.

The non-dispersive imaging path is easily to understand due to only mirror reflection and lens focusing involved. The light intensity will be modulated by the coding aperture (Hadamard-S matrix) on DMD. Taking the 7×7 Hadamard-S matrix as an example, Fig. 2 shows the coding process.

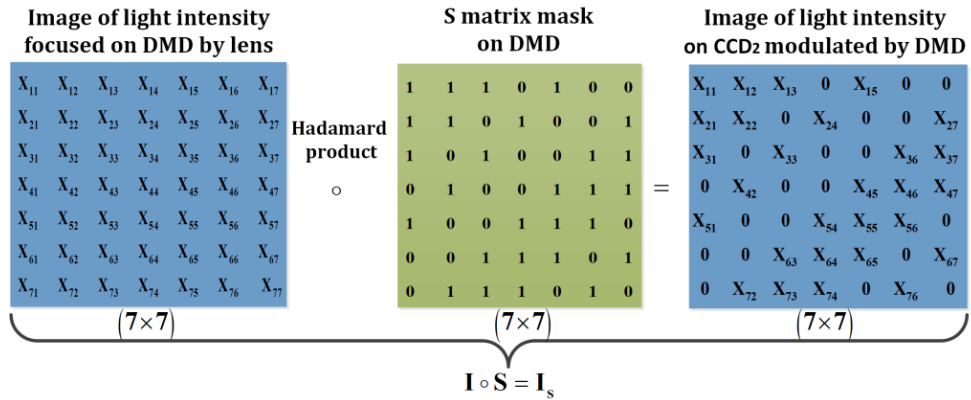

Fig. 2 Coding process in non-dispersive light path.

Through using suitable focal length lens, we can guarantee each pixel of DMD is imaged by one or more pixels of camera. Then, we use the method mentioned in [21] to extra the light intensity of the scene. The implementation of snapshot HTS could be turned into a multiplexing coding spectrometer (Fig. 3), if the non-dispersive imaging path is not considered, described in [4, 16-18].

The measurement problem of dispersive imaging path in traditional HTS[4, 16-18] can be described as:

$$\mathbf{C} = \mathbf{S}(\mathbf{I} \circ \mathbf{f}) \tag{1}$$

where, $\mathbf{S}$ is the coding matrix, $\mathbf{I}$ is the light intensity, $\circ$ is Hadamard product operation, $\mathbf{f}$ is the spectra need to be measured, $\mathbf{C}$ is the measured overlapped spectra. In traditional HTS, the measured spectrum is modulated by the light intensity.

Traditional HTS needs $n$ times measurements to reconstruct the spectrum, obtaining reliable SNR improvement, $n$ is the order of coding matrix. In the snapshot HTS, the light intensity distribution of aperture is modulated by the coding matrix, which can be stated as $\mathbf{I_S} = \mathbf{S} \circ \mathbf{I}$. In the snapshot HTS, mapping the whole coding matrix (Hadamard-S matrix) to the DMD to encodes all spectra. Based on our pervious works[9, 13, 15], we take 7×7 Hadamard-S matrix as an example to explain the physical mechanism and mathematical expression of snapshot HTS in Fig. 4.

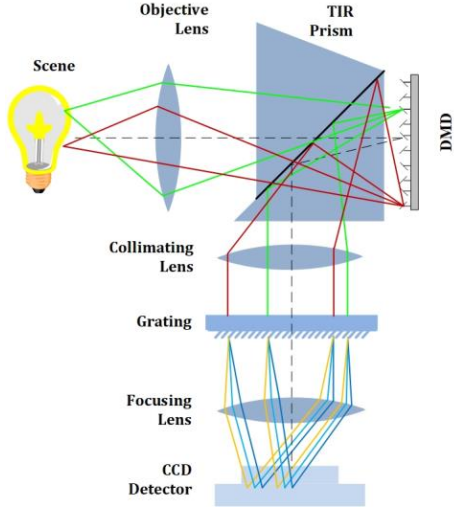

Fig. 3 The dispersive imaging path.

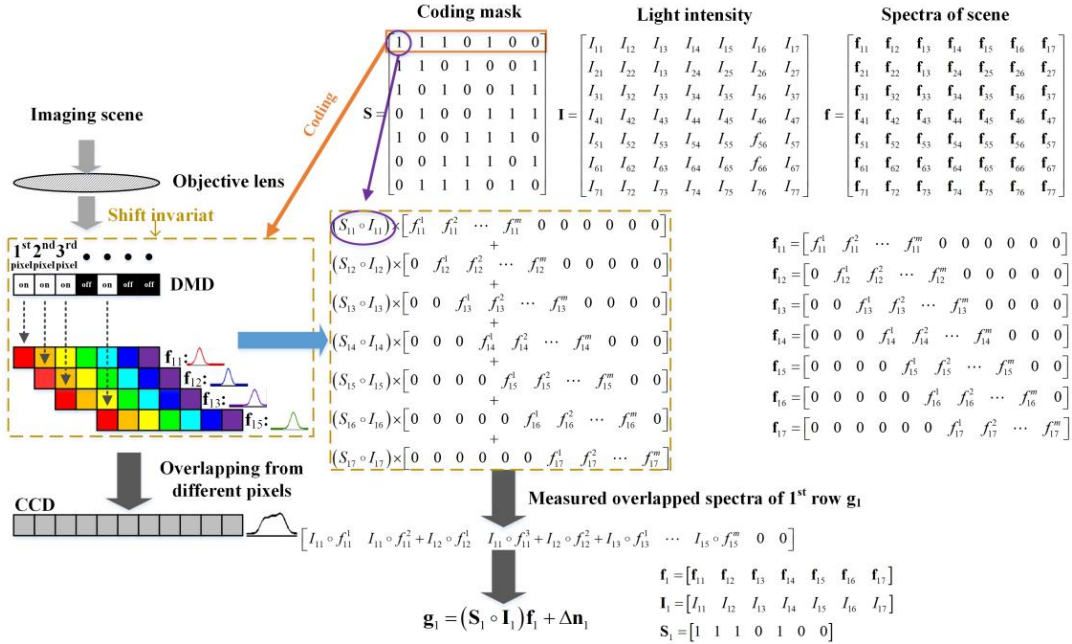

Fig. 4 Physical mechanism and mathematical expression of Hadamard-S matrix coding

In Fig. 4, $\mathbf{S}_1$ is 1$^{st}$ row of Hadamard-S matrix, $\mathbf{I}_1$ is the 1$^{st}$ row light intensity distribution of the scene, $\mathbf{f}_1$ is the spectrum of 1$^{st}$ row of the scene, $\mathbf{f}_{11}$ is the spectrum of 1$^{st}$ pixel in 1$^{st}$ row and $m$ represents its range, the zeros in the spectrum represent the shift invariant and make the processing easy to understand. The measurement problem of $i^{th}$ row of snapshot HTS is stated as:

$$\mathbf{g}_i = (\mathbf{S}_i \circ \mathbf{I}_i)\mathbf{f}_i + \mathbf{n}_{\text{snap}_i} \tag{2}$$

where, $\mathbf{g}_i$ is overlapped dispersed spectra of $i^{th}$ row spectra, $\mathbf{S}_i$ is the $i^{th}$ row of Hadamard-S matrix, $\mathbf{I}_i$ is the $i^{th}$ row of light intensity distribution $\mathbf{I}$, $\mathbf{f}_i$ is the $i^{th}$ row spectra need to be measured, $\mathbf{n}_{\text{snap}_i}$ is measurement noise in $i^{th}$ row measurement. If the light intensity at each element of coding aperture is exactly the same, the normalized intensity of $\mathbf{I_S}$ would turn into a standard Hadamard-S matrix, stated as $\mathbf{S} = \mathbf{I_S}/\max(\mathbf{I_S}) = \mathbf{S}_{\text{snap}}$. Unfortunately, the intensity of light would not be the same, the $\mathbf{S}_{\text{snap}}$ will be approximate the standard Hadamard-S matrix, calling $\mathbf{S}_{\text{snap}}$ as sub-Hadamard-S matrix (sub S-matrix). Specifically, the $\mathbf{S}_{\text{snap}}$ has a key feature, which state as:

$0 \leq \mathbf{S}_{snap}(i,j) \leq 1, \{i \leq n, j \leq n\}$. If each row of the scene has the same spectra or we want to measure the average spectra of each column, i.e. $\mathbf{f}_1 = \mathbf{f}_2 = \cdots = \mathbf{f}_n$ or $\mathbf{f}_{11} = \mathbf{f}_{21} = \cdots = \mathbf{f}_{n1} = \frac{1}{n}\sum_{i=1}^{n}\mathbf{f}_{i1}$. Based on this acceptable assumption, the whole measurement of snapshot HTS can be simplified as:

$$\mathbf{g} = (\mathbf{S} \circ \mathbf{I})\mathbf{f} + \mathbf{n}_{snap} = \mathbf{S}_{snap}\mathbf{f} + \mathbf{n}_{snap} = (\mathbf{S} - \mathbf{S}_1)\mathbf{f} + \mathbf{n}_{snap} \quad (3)$$

where, $\mathbf{S}_1$ is the difference between Hadamard-S matrix $\mathbf{S}$ and the normalized modulated light intensity distribution $\mathbf{S}_{snap}$, which satisfies $0 \leq \mathbf{S}_1(i,j) < 1$.

Unlike NNLS, the problem of snapshot HTS is a well-defined problem, which means that we can obtain a certain and stable SNR boost. In the next section, we will analysis the SNR boost of snapshot HTS more detailed.

## 3. SNR analysis

In order to compare the denoising performance of sub S-matrix with Hadamard-S matrix, firstly, the restored spectrum $\hat{\mathbf{f}}$ can be stated as:

$$\hat{\mathbf{f}} = \mathbf{f} + (\mathbf{S} - \mathbf{S}_1)^{-1}\mathbf{n}_{snap} \quad (4)$$

Then the SNR boost of $(\mathbf{S} - \mathbf{S}_1)$ can be stated as:

$$SNR_{\hat{\mathbf{f}}} = 10\log\left(\frac{\mathbf{f}^T\mathbf{f}}{\mathbf{n}_{snap}^T(\mathbf{S}^T - \mathbf{S}_1^T)^{-1}(\mathbf{S} - \mathbf{S}_1)^{-1}\mathbf{n}_{snap}} \Big/ \frac{\mathbf{f}^T\mathbf{f}}{\mathbf{n}_{snap}^T\mathbf{n}_{snap}}\right)$$

$$= 10\log\left(\frac{\mathbf{n}_{snap}^T\mathbf{n}_{snap}}{\mathbf{n}_{snap}^T(\mathbf{S}^T - \mathbf{S}_1^T)^{-1}(\mathbf{S} - \mathbf{S}_1)^{-1}\mathbf{n}_{snap}}\right) \quad (5)$$

Denote that $\mathbf{n}'_{snap} = (\mathbf{S} - \mathbf{S}_1)^{-1}\mathbf{n}_{snap}$, then Eq. (5) can be rewritten as:

$$SNR_{\hat{\mathbf{f}}} = 10\log\left(\frac{\mathbf{n}_{snap}^T\mathbf{n}_{snap}}{\mathbf{n}_{snap}^T(\mathbf{S}^T - \mathbf{S}_1^T)^{-1}(\mathbf{S} - \mathbf{S}_1)^{-1}\mathbf{n}_{snap}}\right)$$

$$= 10\log\left(\frac{(\mathbf{n}'_{snap})^T(\mathbf{S}^T\mathbf{S} - \mathbf{S}^T\mathbf{S}_1 - \mathbf{S}_1^T\mathbf{S} + \mathbf{S}_1^T\mathbf{S}_1)\mathbf{n}'_{snap}}{(\mathbf{n}'_{snap})^T\mathbf{n}'_{snap}}\right) \quad (6)$$

Suppose that the minimum non-zero value in $\mathbf{S}_{snap}$ is $\alpha$, the larger $\alpha$ is, the more smooth of the light intensity is. Thus $\mathbf{S}_{snap}$ can be rewritten as $\mathbf{S}_{snap} = \alpha\mathbf{S} + \mathbf{S}_2$, where $\mathbf{S}_2$ is the difference between $\alpha\mathbf{S}$ and $\mathbf{S}_{snap}$. Then, we obtain $\mathbf{S} - \mathbf{S}_1 = \alpha\mathbf{S} + \mathbf{S}_2$, $(1-\alpha)\mathbf{S} = \mathbf{S}_1 + \mathbf{S}_2$, $k\mathbf{S} = \mathbf{S}_1 + \mathbf{S}_2$, where $k = 1 - \alpha$, is the disturbance of light intensity at the coding aperture. As we pointed out previously, the $\mathbf{S}_1$ meets $0 \leq \mathbf{S}_1(i,j) < 1$. Thus, $0 \leq \mathbf{S}_2(i,j) \leq k$, $0 \leq k < 1$, and Eq. (6) can be written as:

$$SNR_{\hat{\mathbf{f}}} = 10\log\left(\frac{(\mathbf{n}'_{snap})^T\left((1-k)^2\mathbf{S}^T\mathbf{S} + \frac{1-k}{k}\mathbf{P}\right)\mathbf{n}'_{snap}}{(\mathbf{n}'_{snap})^T\mathbf{n}'_{snap}}\right) \quad (7)$$

According to the Eq. (7), $(\mathbf{n}'_{snap})^T\mathbf{P}\mathbf{n}'_{snap}$ is the key to estimate the $SNR_{\hat{\mathbf{f}}}$. In order to make a further estimation of $(\mathbf{n}'_{snap})^T\mathbf{P}\mathbf{n}'_{snap}$, one hypothetical condition is assumed in this paper, the noise $\mathbf{n}_{snap}$ is independent of random coding matrix $\mathbf{S}_1$, so is the $\mathbf{n}'_{snap}$. Considering $\mathbf{S}_2 = k\mathbf{S} - \mathbf{S}_1$, $\mathbf{S}_1$ and $\mathbf{S}_2$ follow the same distribution, so $\mathbf{n}_{snap}$ will be independent of $\mathbf{S}_2$ too. In brief, following conclusions can be inferred.

$$\operatorname{cov}(\mathbf{S}_1(1,j), \mathbf{n}'_{snap}) \approx 0$$

$$\sum_{j=1}^{n}\left(\mathbf{S}_1(1,j) \times \mathbf{n}'_{snap}(j)\right) \propto E(\mathbf{S}_1(1,j)) \times E(\mathbf{n}'_{snap}(j)) \quad (8)$$

$$(\mathbf{n}'_{snap})^T\mathbf{S}_1^T\mathbf{S}_1\mathbf{n}'_{snap} \propto \sum_{k=1}^{n}\left\{E(\mathbf{S}_1(k,j)) \times E(\mathbf{n}'_{snap}(j))\right\}^2$$

Since $\mathbf{S}_1$ and $\mathbf{S}_2$ follow the same distribution, so the mean of $\mathbf{S}_1(k,x)$ and $\mathbf{S}_2(k,x)$ will be approximate. It means that following equation can be obtained:

$$E(\mathbf{S}_1(k,x)) \approx E(\mathbf{S}_2(k,x)) \quad (9)$$

Based on Eq. (8) and (9), $(\mathbf{n}'_{snap})^T\mathbf{S}_1^T\mathbf{S}_1\mathbf{n}'_{snap}$ and $(\mathbf{n}'_{snap})^T\mathbf{S}_2^T\mathbf{S}_2\mathbf{n}'_{snap}$ will be extremely approximate. It means that following equation is confidential, which can be stated as:

$$(\mathbf{n}'_{snap})^T\mathbf{S}_1^T\mathbf{S}_1\mathbf{n}'_{snap} \approx (\mathbf{n}'_{snap})^T\mathbf{S}_2^T\mathbf{S}_2\mathbf{n}'_{snap} \quad (10)$$

Considering $2 \leq (2-k)/(1-k)$, based on Eq. (10), $\mathbf{P}$ can be rewritten as:

$$\mathbf{P} = \mathbf{S}_1^T\mathbf{S}_2 + \mathbf{S}_2^T\mathbf{S}_1 + (2-k)/(1-k)\mathbf{S}_2^T\mathbf{S}_2$$

$$\approx \mathbf{S}_1^T\mathbf{S}_1 + \mathbf{S}_1^T\mathbf{S}_2 + \mathbf{S}_2^T\mathbf{S}_1 + 1/(1-k)\mathbf{S}_2^T\mathbf{S}_2 \quad (11)$$

$$\geq (\mathbf{S}_1 + \mathbf{S}_2)^T(\mathbf{S}_1 + \mathbf{S}_2) = k^2\mathbf{S}^T\mathbf{S}$$

Based on Eq. (6) and (11), $SNR_{\hat{\mathbf{f}}}$ can be simplified as:

$$SNR_{\hat{\mathbf{f}}} \geq 10\log\left(\frac{(\mathbf{n}'_{snap})^T\left((1-k)^2\mathbf{S}^T\mathbf{S} + \frac{1-k}{k}k^2\mathbf{S}^T\mathbf{S}\right)\mathbf{n}'_{snap}}{(\mathbf{n}'_{snap})^T\mathbf{n}'_{snap}}\right)$$

$$= 10\log\left(\frac{(\mathbf{n}'_{snap})^T\mathbf{S}^T\mathbf{S}\mathbf{n}'_{snap}}{(\mathbf{n}'_{snap})^T\mathbf{n}'_{snap}}\right) + 10\log(1-k) \quad (12)$$

Furthermore, we evaluated the denoising capability of sub-Hadamard-S matrix $(\mathbf{S}-\mathbf{S}_1)$ in simulation (see in Fig. 5). Fortunately, the $(\mathbf{S}-\mathbf{S}_1)$ works well in the simulation. The SNR of snapshot HTS degrades about $10\log\left(\frac{1}{1-k}\right)$ (dB) compared with traditional HTS, which meets the conclusion obtained in Eq. (12). In snapshot HTS, the ground glass will make the disturbance factor lower than 0.5, in this way snapshot HTS will obtain more obvious SNR boost.

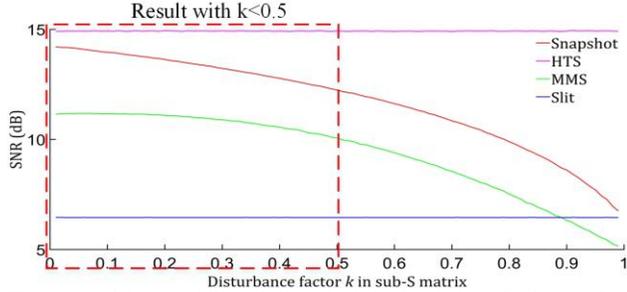

Fig. 5. Denoising capability of all methods in simulation with *k* along 0 to 0.99.

## 4. Simulation and Experimental Results

In order to demonstrate the denoising capability of sub-Hadamard-S matrix, we benchmark snapshot HTS in both simulations and experiments. Fig. 5 shows the denoising capability comparisons among snapshot HTS, HTS, MMS and slit-based spectrometer among the disturbance of light intensity *k* from 0.01 to 0.99 as $0.01, 0.02, \cdots, 0.99$. Specifically, $127\times127$ Hadamard-S matrix is employed, the normalized solar spectrum is utilized as the original reflectance data, White Gaussian noise is simulated as detector noise. For the robustness, we perform all of the instances 100 times, and present the average value as the final result.

Since the NNLS method belong to so-called ill-posed inverse problem, in MMS, when $127\times127$ Hadamard-S matrix is employed, we will obtain 127 reconstructed spectra. However, in these spectra, it is hard to seek the "best" or the highest SNR spectrum, since the SNR of these spectra are quite different. In the simulation results, we record all SNR of reconstructed spectra in each simulation and we use mean SNR boost of each spectrum and the confidence interval of all measured spectra to evaluate the stable of all method.

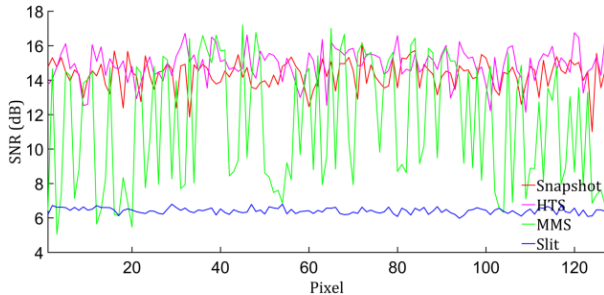

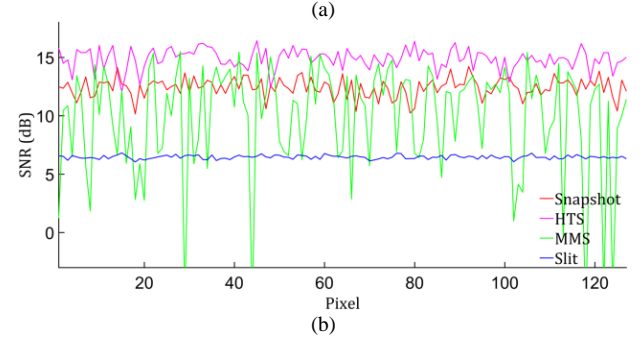

Fig. 6. SNR distribution of 127 reconstructed spectra of all methods. (a) Disturbance factor of light distribution *k*=0.1 (b) *k*=0.5.

As shown in Fig. 6, the SNR boost of spectra located in different rows reconstructed by NNLS, are quite different. With the disturbance of light intensity increase, the SNR distribution will become more and more unstable. In practical, for a given reconstructed spectrum in MMS, we cannot ensure its SNR boost and limits its applications. However, snapshot HTS and traditional HTS have much more stable SNR boost.

In order to evaluate the distribution of SNR of all methods, we employ confidence interval to measure stableness of all methods based on the $100\times127=12700$ simulated spectra. Suppose that the distribution of SNR obey Gaussian distribution with mean value $\mu$ and variance $\sigma^2$. According to law of large numbers and central limit law, the simulated results obey Gaussian distribution with mean value $\mu$ and variance $\sigma^2/12700$. Thus, in 95% confidence interval, the range of different methods are shown in Table. 1.

Table. 1 Confidence interval of all methods.

|  | *k*=0.1 | *k*=0.5 |
|---|---|---|
| Snapshot | 12.52-15.87 | 10.79-14.04 |
| HTS | 13.20-16.62 | 13.23-16.63 |
| MMS | 4.06-18.94 | 3.04-17.66 |
| Slit | 6.14-6.77 | 6.14-6.76 |

As shown in the Table. 1, for a given spectrum measured by MMS, its SNR distribution range is too wide, and with the disturbance of light intensity increase the distribution range will become more and more wide. That is to say, the lower bound of SNR boost is too small even smaller than slit-based spectrometer when the disturbance of light intensity is obvious. The SNR distribution range of snapshot HTS and HTS is much narrower than MMS, which means the lower bound of SNR boost of snapshot HTS and HTS is much higher than MMS. Slit-based spectrometer has the narrowest SNR distribution range, however, its SNR boost is too small and cannot be used to measure weak intensity signal.

We show some representative reconstructed spectra of all methods with the disturbance factor $k=0.1, 0.5$ in Fig. 7, respectively. For MMS, we select the spectrum with the mean SNR of all 127 reconstructed spectra.

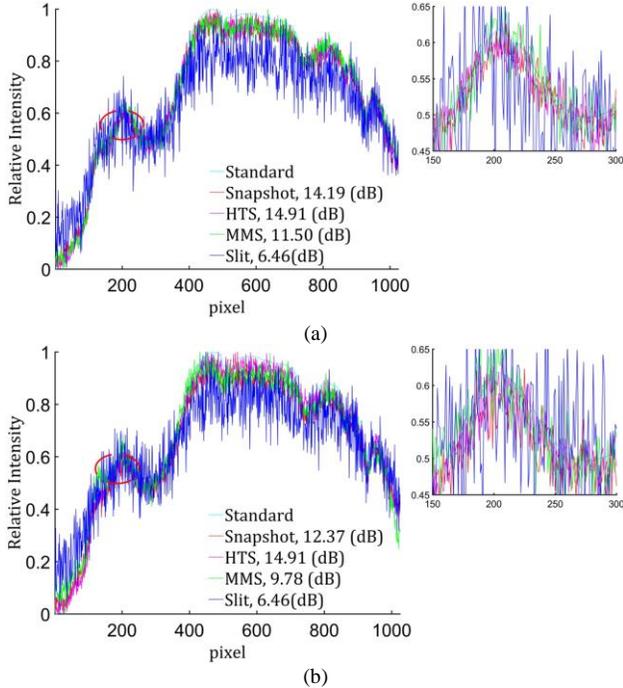

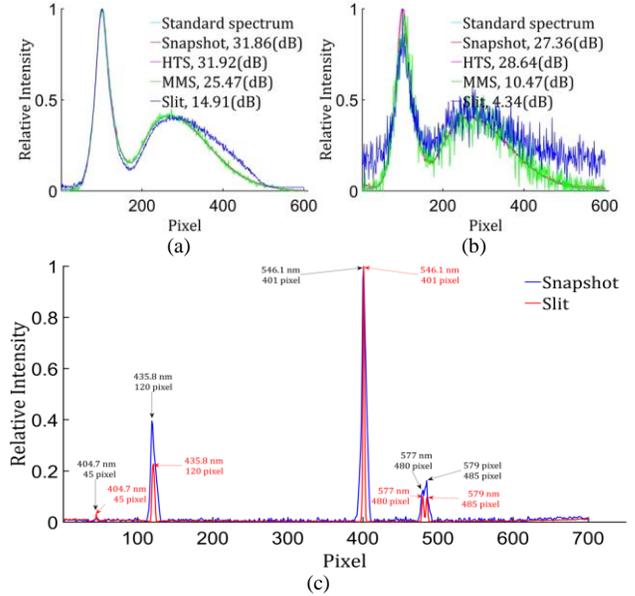

Fig. 7. Simulation results of denoising capability of the s snapshot HTS, HTS, MMS and slit-based spectrometer while detector noise dominated. (a) Disturbance factor of light distribution $k=0.1$ (b) $k=0.5$.

Fig. 8. Experimental results with snapshot HTS, HTS, MMS and slit-based spectrometer. (a) Experimental results with 10ms and 0dB. (b) Experimental results with 100us and 36dB. (c) Spectrum of Hg lamp measured by snapshot HTS and slit-based spectrometer.

According to the simulation results (Fig. 7), snapshot HTS and HTS significantly outperform slit-based spectrometer, while detector noise dominated. However, we can find that, the advantage of snapshot HTS would decrease along with the increase of disturbance of light. When the light distribution of the scene is very smooth, HTS and snapshot HTS have almost the same performance. Though the variation of scene would degrade the denoising performance of snapshot HTS, it still has a significant improvement of SNR. Like snapshot HTS, the performance of MMS would decrease along with the increase of disturbance of light. However, as mentioned above, the SNR boost of MMS is not stable and limits its applications.

To evaluate the denoising capability in the real scene, we measure the SNR of snapshot HTS, HTS, MMS and slit-based spectrometer in low level light. In the experiment, we use an absorptive neutral density filter to cover an LED to simulate the low level light source. We use slit-based spectrometer to obtain the standard spectrum with high light of LED and long exposure time. In order to compare performance of all methods quantitatively, we set the same exposure time in snapshot HTS, HTS, MMS and slit-based spectrometer. Since we have the standard spectrum, we select the highest SNR of reconstructed spectra of MMS as the result of MMS. The final results are shown in Fig. 8.

In Fig. 8, 10ms and 0dB means the exposure time of CCD is 10ms and gain is 10dB, so is 100 μs and 36dB. The experimental results have demonstrated that compared to slit-based spectrometer, snapshot HTS has much higher sensitivity and much better robustness (Fig. 8(a), (b)). Compared to MMS, snapshot HTS is a well-defined problem and can obtain better results. Compared to HTS, snapshot HTS has almost the same performance. That is to say, compared to the standard Hadamard-S matrix, the sub-Hadamard-S matrix has almost the same denoising capability.

Moreover, Hg lamp is employed to obtain the spectral resolution of snapshot HTS. The result of snapshot HTS is shown in Fig. 8(c). The corresponding peaks (577 nm and 579 nm) are located at 480 pixel and 485 pixel, that is to say, the spectral resolution of snapshot HTS is higher than 1 nm/pixel. Compared to traditional silt-based spectrometer, the proposed snapshot HTS has almost the same spectral resolution.

## 5. Conclusion

In brief, we proposed a snapshot HTS based on sub-Hadamard-S matrix in this paper. As far as we know, it is the first time a coding spectrometer is realized, having comparable SNR boost with HTS, maintaining snapshot advantage. It demonstrated that sub-Hadamard-S matrix coding still has significant SNR boost. The denoising capability of coding matrix is quite robust while the element of coding matrix is varying. This feature makes the high SNR snapshot spectrometer feasible. Different from traditional HTS, an extra imaging path to collect the non-dispersed image of coding aperture is added in the proposed system. Therefore, we could turn the Hadamard-S matrix coding measurement into a sub-Hadamard-S matrix coding. Though the extra imaging path increase the complexity of the system, we can realize snapshot and obtain stable SNR boost. More importantly, different from snapshot spectrometer with convolution or compressive sensing, it is a kind of well-defined problem, which means the SNR boost has a lower bound. Compared to slit-based spectrometer, snapshot HTS could gain much higher SNR. Compared to traditional HTS, snapshot HTS can achieve comparable performance but with much higher speed. Compared to numerical method such as, MMS and dual-camera coded aperture snapshot spectral imaging, snapshot HTS is a well-defined problem and can obtain better and more stable results.

**Acknowledgements** This work was supported by the National Natural Science Foundation of China (Grant Nos. 61727802, 61601225). We thank Enlai Guo, Zhaoxin Wang for technical supports.

## References

1. M. E. Gehm, R. John, D. J. Brady, R. M. Willett and T. J. Schulz, "Single-shot compressive spectral imaging with a dual-disperser architecture", Opt. Express 15 (2007) 14013-14027.
2. C. V. Correa, H. Arguello and G. R. Arce, "Snapshot colored compressive spectral imager", J. Opt. Soc. Am. A 32 (2015) 1754-1763.
3. A. Mrozack, D. L. Marks and D. J. Brady, "Coded aperture spectroscopy with denoising through sparsity", Opt. Express 20 (2012) 2297-2309.
4. A. Wagadarikar, R. John, R. Willett and D. J. Brady, "Single disperser design for coded aperture snapshot spectral imaging", Appl. Opt. 47 (2008) 44-51.
5. D. Kittle, K. Choi, A. Wagadarikar and D. J. Brady, "Multiframe image estimation for coded aperture snapshot spectral imagers", Appl. Opt. 49(36) (2010) 6824-6833.
6. Y. Wu, I. O. Mirza, G. R. Arce and W. P. Dennis, "Development of a digital-micromirror-device-based multishot snapshot spectral imaging system", Opt. Lett. 36(14) (2011) 2692-2694.
7. F. Kazemzadeh and A. Wong, "Resolution- and throughput-enhanced spectroscopy using a high-throughput computational slit", Opt. Lett. 41 (2016) 4352-4355.
8. X. Ma, H. Wang, Y. Wang, D. Chen, W. Chen and Q. Li. "Improving the resolution and the throughput of spectrometers by a digital projection slit", Opt. Express, 25(19) (2017) 23045-23050.
9. J. Yue, J. Han, Y. Zhang and L. Bai, "High-throughput deconvolution-resolved computational spectrometer", Chin. Opt. Lett. 12(4) (2014) 043001.
10. M. Harwit and N. J. A. Sloane, Hadamard Transform Optics (Academic Press, 1979).
11. M. J. E. Golay, "Multi-Slit Spectrometry*", J. Opt. Soc. Am. 39 (1949) 437-444.
12. M. Chi, Y. Wu, F. Qian, H. Peng, W. Zhou and Y. Liu, "Signal-to-noise ratio enhancement of a Hadamard transform spectrometer using a two-dimensional slit-array", Appl. Opt. 56(25) (2017) 7188-7193.
13. J. Yue, J. Han, Y. Zhang and L. Bai, "Denoising analysis of Hadamard transform spectrometry", Opt. Lett. 39 (2014) 3744-3747.
14. J. A. Decker, "Experimental Realization of the Multiplex Advantage with a Hadamard-Transform Spectrometer", Appl. Opt. 10 (1971) 510-514.
15. J. Yue, J. Han, L. Li and L. Bai, "Denoising analysis of spatial pixel multiplex coded spectrometer with Hadamard H-matrix", OPT. COMMUN. 407 (2018) 355-360.
16. A. A. Wagadarikar, M. E. Gehm and D. J. Brady, "Performance comparison of aperture codes for multimodal, multiplex spectroscopy", Appl. Opt. 46(22) (2007) 4932-4942.
17. M. E. Gehm, S. T. McCain, N. P. Pitsianis, D. J. Brady, P. Potuluri, and M. E. Sullivan, "Static two-dimensional aperture coding for multimodal, multiplex spectroscopy," Appl. Opt. 45 (2006) 2965.
18. C. Fernandez, B. D. Guenther, M. E. Gehm, D. J. Brady, and M. E. Sullivan, "Longwave infrared (LWIR) coded aperture dispersive spectrometer," Opt. Express, 15(9) (2007) 5742-5753.
19. L. Wang, Z. Xiong, D. Gao, G. Shi, F. Wu, "Dual-camera design for coded aperture snapshot spectral imaging", Appl. Opt. 54(4) (2015) 848-858.
20. J. M. Bioucas Dias, M. A. T. Figueiredo, "A new TwIST: Two-step iterative shrinkage/thresholding algorithms for image restoration", IEEE T IMAGE PROCESS 16(12) (2007) 2992-3004.
21. L. Gao, R. T. Kester, N. Hagen, T. S. Tkaczyk, "Snapshot image mapping spectrometer (IMS) with high sampling density for hyperspectral microscopy", Opt. Express, 18(14) (2010): 14330-14344.